\documentclass[referee]{aa} % for a referee version
\input psfig.sty

\input{epsf.tex}

\newcommand{\ltapprox}{\raisebox{-0.5ex}{$\,\stackrel{<}{\scriptstyle\sim}\,$}}
\newcommand{\gtapprox}{\raisebox{-0.5ex}{$\,\stackrel{>}{\scriptstyle\sim}\,$}}

\begin{document}
 
\thesaurus{03(11.05.1; 11.19.2; 11.12.2)}
% 11.03.4 Coma (=Abell 1656);
%
\title{Extensive near-infrared ($H$-band) photometry in Coma 
\thanks{Based on observations collected with the {\it T\'elescope 
Bernard Lyot}, at the Pic du Midi Observatory, operated by INSU (CNRS)}}
 
\author{S. Andreon@1,  R. Pell\'o@2,  E. Davoust@2, R. Dom\' \i nguez@2@3, 
   P. Poulain@2}
 
%   \offprints{S. Andreon}
 
\institute{
   @1
  Osservatorio Astronomico di Capodimonte, via Moiariello 16,
	80131 Napoli, Italy (email: andreon@na.astro.it)  
\and
   @2
   Observatoire Midi-Pyr\'en\'ees, LAT, UMR 5572,
   14 Avenue E. Belin, F-31400 Toulouse, France 
   France (email: roser,davoust,poulain@obs-mip.fr)
\and
   @3
   Departament d'Astronomia i Meteorologia, Marti i Franqu\`es 1, 08028 Barcelona,
        Spain
}
 
\date{Received, 1999, ... Accepted, ...}

\authorrunning{Andreon et al.}
\titlerunning{$H$ photometry in Coma} 

\maketitle
 
\begin{abstract}

We present extensive and accurate photometry in the near-infrared $H$ band of 
a complete sample of objects in an area of about 400 arcmin$^2$ toward the 
Coma cluster of galaxies. The sample, including about 300 objects, is 
complete down to $H\sim17$ mag, the exact value depending on the type of 
magnitude (isophotal, aperture, Kron) and the particular region studied.  This 
is six magnitudes below the characteristic magnitude of galaxies, well into the 
dwarfs' regime at the distance of the Coma cluster. 
For each object (star or galaxy) we provide aperture 
magnitudes computed within five different apertures, the magnitude within the 
22 mag arcsec$^{-2}$ isophote, the Kron magnitude and radius, magnitude 
errors, as well as the coordinates, the isophotal area, and a stellarity 
index. Photometric errors are 0.2 mag at the completness limit.
This sample is meant to be the zero-redshift reference for evolutionary 
studies of galaxies. 

\end{abstract}

\section{Introduction}

Our present knowledge of the near-infrared properties of galaxies is still 
very sketchy, because panoramic
astronomical imaging receivers in that domain have only 
been available recently.  For samples complete in the near-infrared band, 
properties such as galaxian colors and 
sizes in the near-infrared and their dependence on morphological type are 
almost unknown.  And so are the deep near-infrared luminosity functions of 
galaxies in the field and in clusters.  The former have been investigated by 
Gardner et al. (1997) and, for an optically-selected sample, by Szokoly et 
al. (1998),  but they cover just its bright end.  The latter concern the 
bright end of intermediate-distance clusters (Trentham \& Mobasher 1998; 
Barger et al. 1996); an exception is the study of the luminosity function of 
the center of Coma by de Propris et al. (1998), which reaches $H \sim 16$ mag
(at 80\% completeness).

Near-infrared properties of galaxies are very useful for a number of 
projects.  These quantities allow the determination of a possible change with
redshift in 
the galaxy properties, such as the ``downsizing" suggested by Cowie et al.
(1996). 
The passband dependence, from the optical to the near-infrared, of the tilt 
of the fundamental plane gives a clear indication of whether this tilt is due 
to a variation of metallicity, age or a breakdown in homology along the 
early-type sequence (Pahre et al. 1998), but the 
fundamental plane in the near-infrared is only beginning to be studied. 
The existence of dwarf galaxies with blue $B-K$ colors has been assumed, in 
order to explain galaxy counts in $B$ and $K$ (Saracco et al.
1996), but their density in the universe still has to be measured. 

Near-infrared measurement of nearby galaxies and clusters are valuable, 
because even at high redshift the near-infrared falls into a well 
known part of the restframe spectrum, at the difference of the optical window, 
which, at high redshift, samples the almost unexplored restframe 
ultraviolet emission. 
In passing, the galaxy size distribution plays a fundamental role in 
performing new cosmological tests (Petrosian, 1998). Their determination in 
the present Universe is critical for these tests. 

In short, good reference samples of galaxies with well established 
spectrophotometric properties -- from the optical to the near-infrared -- are 
urgently needed for the study of galaxy properties and their evolution.

% fig 1.

We thus started a program to establish the near-infrared photometric 
properties of galaxies in Coma, the archetypical zero-redshift cluster, for 
which a large amount of reliable data (photometry {\it and} morphological 
types) are already available.  We adopted the $H$ band because it corresponds 
to the $K$ band -- still observable from the ground -- for galaxies at 
intermediate redshift (z $\sim$ 0.3), thus allowing a direct comparison of the 
galaxy properties over a redshift range where evolution in clusters has been 
detected (Butcher \& Oemler 1984). $J$--band ($1.2 \mu$) images for
the same region have been already aquired and their reduction is in 
progress.

\section{Strategy of observation}

The studied region is $\sim20 \times 24$ arcmin wide, located $\sim15$ arcmin North-East 
from the Coma cluster center (see Fig. 1). The approximate center of the 
region is 13~00~35 +28~14~35 (J2000). We voluntarily avoided imaging the Coma 
cluster core, because of the possible presence of a halo around the two 
dominant galaxies, which would make the reduction and analysis of the data 
more difficult, given the small field of view of our camera. 

Data were taken in the near-infrared $H$ band, on March 4th and 9th, and 
April 3rd to 5th, 1997, at the 2-meter {\it T\'elescope Bernard Lyot} 
of Pic du 
Midi Observatory (France), with the {\it Mo\" \i cam} camera set at its f/8 
focus. {\it Mo\" \i cam} is a $256\times256$ pixel camera, based on an array of $2 
\times 2$ NICMOS-3 detectors. The adopted configuration gives the maximum field 
of view available, 2' $\times$ 2', with a pixel size of 0.5 arcsec. A few 
hours per night were allocated to us in March, and half-nights in April. 

For an efficient use of the assigned observing time, we did not guide the 
telescope during the exposures. This limits the individual exposure times to 
30 seconds, without appreciable degradation of the PSF. This is also an optimum 
value for the exposure time in $H$, in order to reach a suitable 
signal-to-noise ratio on individual exposures, while keeping the sky 
background relatively low ($\sim3000$ to $6000$ ADU). In order to image the 
$20 \times 24$ arcmin region with the camera, we mosaicked the total field by 
successive strips in declination, moving the telescope by half a field of view 
in declination between exposures, and by half a field of view in right 
ascension between strips. In this way, with the exception of the regions close 
to the edge of each
field, each point in the sky is observed four times, giving a 
total exposure time per sky pixel of 120 seconds.  

Almost half of the field was observed more than four times in order to reach 
fainter magnitudes.  This was achie\-ved in three ways: by doubling the number 
of exposures at each telescope pointing, by moving the telescope by 1/3 of 
a field of view instead of 1/2, or by observing the same strip on different 
nights. The final median exposure times are 310, 120 and 400 seconds 
respectively for the fields patch1 (0403+0903+0304), 0404 and 0504, as were 
dubbed the regions -- from their date of observation.

% fig 2

Unfortunately, the telescope does not point with a precision of one pixel, and 
therefore, our 675 manually introduced telescope offsets\footnote{One every 30 
seconds; we warmly thank the telescope operators for the work involved} are 
only approximately achieved on the sky (see Fig. 2).  These errors 
accumulate to create a field which is not exposed in a perfectly uniform way, 
and, since the actual offset is a few arcsec smaller than the required one and 
systematic errors accumulate in RA, there are regions exposed more than 
required and others never exposed. This problem was unknown at the time of the 
observations, probably because no-one had tried to use our observational 
strategy before at this telescope (we observed during the very first observing 
runs with the NICMOS-3 camera). However, since we returned repeatedly 
to the same regions, 
only one narrow gap is left in the whole field, between fields 0504 and 0404.  
Two bright galaxies of Coma fall in this gap, GMP 2413 and GMP 2418, as well as 
several QSOs and stars. 

All images have been taken with the same exposure time, 30 seconds, in order to
simplify the data reduction.

\section{Data reduction}

\subsection{Dark}

The $masterdark$ is the median of a stack of 25 individual darks. The rms
noise of an individual dark is about 16 to 23 ADU, depending on the NICMOS-3
quadrant. In the $masterdark$, pixels whose intensity differs by more than 70
ADU from the surrounding ones (they are $\sim1$\% of the total number of pixels)
were replaced by the local average. These pixels 
were found to show a different behavior from their neighbors at a low 
illuminating flux, and a normal behaviour at high illuminating flux
(i.e. in science and calibration images).

The gain of the camera was found to be 10 e$^-$/ADU.

\subsection{Flatfielding}

This is a challenging task with near-infrared images, and therefore a 
detailed description of the way we flatfielded our images is necessary.  In 
the near-infrared, observers are faced with rapid changes of the intensity 
{\it and} spectrum of the sky. In the $H$ band, the measured contribution 
of the telescope and warm-optics background is 80 ADU during the day (in 30 
seconds, shutter closed), and probably less during the night; this is 
negligible with respect to the sky level and total background emissivity 
(shutter open), which is usually $\sim$ 3000 ADU in 30 seconds. We can 
therefore apply an ``optical reduction" to our $H$-band data. 

Our images are quite empty: most pixels are images of the sky most of the 
time, and the shift between successive exposures is larger than the typical 
size of objects. Therefore, we decided not to take offset images, and to use 
the images themselves for flatfielding, thus doubling the time allocated to 
science frames. The procedure to flatfield our near-infrared images is 
basically the same as that used for super-flats in the case of optical images. 

$final = (raw-masterdark)/(superflat-masterdark)$

For each image, the $superflat$ is computed as follows:

a) we consider only images taken not longer before or after the exposure to
be flat--fielded

b) the median of each image is computed

c) we retain only those having a median which differs by less than 5\% from 
that of the image to be flatfielded, with a maximum of 25 images. More liberal 
limits do not increase the quality of the flatfielding as judged by the 
flatness of the flattened image. We verified that this choice gives better 
flatfielded images than taking the temporally nearest 25 (or any number 
between 9 and 25) images for the creation of the superflat. We interpret the 
above result as evidence that the spectrum and the intensity of the sky in the 
near-infrared are correlated.  For more than 2/3 of the images, 25 images whose 
intensities differ by at most 3\% were used. 

d) the images are divided by their median 

e) the superflat is a median, over the stack, of the median normalized images

\subsection{Check of the photometric accuracy of the flatfielding}

We checked the correctness of the flatfielding by comparing the apparent
flux of photometric standards (and of other observed objects) imaged in 
the four quadrants (and at different locations in each quadrant). 0.01 mag is 
the present upper limit of the error determination of any trend between 
location in the field of view and flux. The global photometric quality of the 
data is discussed in Sect. 3.7. 

\subsection{List of bad pixels}

We manually created the bad-pixel list. In general, {\it Mo\" \i cam}  has 
an excellent cosmetics, and only a few regions required to be masked:
two small ($\sim10$-pixel wide) clusters and half a row between two
quadrants.

% fig 3

\subsection{Sky subtraction}

From each flatfielded image we subtracted a constant, the sigma clipped mean 
computed over the whole image. 

In passing, we note that the average $H$-band sky brightness during our 5
nights was 14.2 mag arcsec$^{-2}$, more than one magnitude fainter than 
the values typically encountered at other European northern observatories,
such as the William Herschel Telescope at La Palma, or at TIRGO on Gornergrat. 
It is only 0.1 mag brighter than the mean value observed on Mauna Kea.

\subsection{Spatial offset}

Most images contain at least one bright object which is common to 
other neighboring images. The angular shifts between the 675 pointings have 
been computed manually for several reasons. There was no 
information written on the image-file headers about the position of the 
telescope when the images were acquired. Besides, when no bright objects are 
present, allowing one to produce an accurate position of the centroids 
automatically and blindly, the image have to be recentered manually. This is 
also the only way to correct for the presence of ghosts and remnant images 
(arising from extremely bright sources). Fortunately, these events were rare 
and they were perfectly corrected after the processing. 

\subsection{Photometric calibration}

Standard stars from the UKIRT faint standard list (Casali \& Hawarden 1992) 
were observed once or twice per night. Each standard star was usually exposed 
four times or more, at least once per detector quadrant. Figure 3 shows the 
zero points (solid dots). We supplementary these a few measurements by using
as standards the Coma galaxies whose 
magnitudes are listed in Recillas-Cruz et al. (1990). The latter magnitudes
are taken with an InSb photometer within a 14.8 arcsec diaphragm and have an average
error of 0.02 mag. Of course, in our comparison with Recillas-Cruz et al. (1990)
we use 14.8 arcsec aperture magnitude. These zero--point are shows in Figure 3
as crosses. When stars and galaxies zero--points are 
available, the agreement is good, as also shows in Table 1.

\begin{table}
\caption{Photometric zero point, outside atmosphere, for a 1-sec exposure time.
The number of objects is given in brackets}
\begin{tabular}{lccc}
\hline
\small
Night & Standard & Standard & Adopted \\
& stars & galaxies & zero point\\
\hline
4/3 & 20.76~(1) & 20.69~(3) & 20.76 \\
9/3 &  & 20.76~(2) & 20.76 \\
3/4 &  & 20.76~(2) & 20.76 \\
4/4 & 20.71~(2) & 20.71~(9) & 20.71 \\
5/4 & 20.67~(2) &  & 20.67 \\
\hline 
\end{tabular}
\end{table}

The field imaged on April 3rd largely overlaps with the field imaged on March 
4th and 9th. The magnitudes of the 9 bright objects in the overlapping region 
are in good agreement, with a typical scatter of $0.03$ mag and an average 
difference of only $0.02$ mag, which confirms that these nights were 
photometric. The fields observed the
other nights have too small overlaps to allow a useful 
comparison in the same way. Nevertheless, all along the recentering procedure, 
we have checked the consistency of the photometry between different 
neighboring images of the same bright sources.

Coma was observed through airmasses which differ at most by 0.1 
mag. Assuming an atmospheric extinction of 0.059 mag/airmass, the zero-point 
variation due to airmass variations is at most 0.0059 mag. We neglected this 
variation (and we assumed an airmass of 1.1 for all observations). 

Overall, the agreement between zero points computed from different 
standard objects (UKIRT stars and Recillas--Cruz et al. (1990) galaxies), the 
agreement on the magnitude of the same object observed in different nights
and the consistency of the photometry of the same object in the various individual
30s exposures, all these facts suggest us that
the zero point error in our photometry is very unlikely to be
systematicly worse or variable more than 0.02 mag.

% fig 4.

%%F4

\subsection{Mosaicking}

%In getting telescope time for realizing a wide mosaic by mapping the sky 
%with a small-field camera, we committ ourself to perform the time--comsuming
%task of combining images taken with significantly different 
%pointings on the sky, although this moral obligation is sometime disregarded
%(e.g. by EIS team).

%We are not the first ones to have realized a wide mosaic by mapping the sky 
%with a small-field camera. But we have the merit of having successfully 
%tackled the problem of combining images taken with significantly different 
%pointings on the sky. 

The images were mosaicked using the task {\tt imcombine} under IRAF.  Files 
containing the flux scaling, the background value, and the relative offsets 
were given in input. The task gives in output the composite image, the image 
of the measured dispersion among the input images, the number of pixels used 
(i.e. the exposure map in units of 30 seconds). The expected variance of the 
composite image has been computed from these quantities, assuming standard 
laws for the propagation of errors. 

The task {\tt imcombine} has a number of tunable parameters whose setting 
affects the results. Since we split the desired exposure time into shorter 
exposures, we composited the images by averaging the fluxes measured in the 
different images, as we would have done if ideal conditions (low sky flux, 
perfect NICMOS-3 cosmetics, good telescope tracking) were met. We 
allowed the exclusion of up to 2 values for each sky direction in order to 
remove cosmic rays and transient hot pixels. We verified that other schemes 
(such as taking a median, or a sigma clipped average possibly centered on the 
median, etc.) do not conserve the flux of the objects within our images (many 
objects are sufficiently exposed, even when for only 30 seconds, for 
accurately measuring their magnitudes and comparing them with those measured 
on mosaicked images). 
The reason is that the average of a distribution
has the useful characteristic, by definition,
to be equal to the integral of the distribution divided by the number of data points.
The median, or any other elaborate means of computing
the central location of a distribution (in particular when it is formed by 
a small number of points, as our one) does not have such property
and therefore does not conserve the integral of the distribution
(the total flux), which is the scientific interesting quantity.

In the mosaicking, we started by compositing the images on a night by
night basis. We checked the relative photometry and astrometry first between
the composite image and the parent single 30-sec images, then among nights
when possible. This careful check allowed us to detect mistakes and
systematic errors: displacement errors appear in regions of particularly high
sigma, when the error is small, or as regions of under-average exposure
time with respect to those of the same RA, when errors are large. Photometric
errors (due for example to an incorrect setting of the {\tt imcombine} task),
appear quite clearly, for example as a good agreement between fluxes of
the same object imaged in different 30-sec exposures, but as a discrepant 
flux in the composite image (or in the image of another night). 

In the final image, the seeing is $\sim 1.7$ arcsec FWHM, of which
we estimate that $\sim 0.5$ arcsec have been added during the data
reduction, and with variations (0.5 arcsec) from region to region.

\subsection{Astrometry}

In order to provide accurate positions for the objects in our three
fields, we used
the astrometric catalog of the USNO version 1 (Monet et al. 1999). In the
three composite near-infrared images there are significant ($>2$ arcsec) 
astrometric distortions (i.e. significant deviations from a linear relation
between relative positions measured on our composite images and the actual
positions on the sky). Therefore, we
first established an approximate conversion from the positions of the
objects in our three fields to the USNO using the task {\tt geomap} in
IRAF. We projected our field (represented as in Fig. 7), with the appropriate 
shifts in RA and DEC, on a red image of the field, taken from the
Second Digitized Sky Survey (hereafter POSS-II). Then, we associated
objects having similar sizes, ellipticities, position angles and sky positions in
our Fig. 7 and on the POSS-II. Most near-infrared objects fall within
2 arcsec of their optical counterpart. Finally, we adopted as
coordinates for the near-infrared objects the POSS-II coordinates of the
corresponding optical objects (which correspond to epoch
1993.4, date of the POSS-II plate) and precessed them to J2000
using the USNO catalog. The final accuracy of the absolute position is 1
arcsec, given by the quadratic sum of the accuracy of the USNO and the
scatter observed in the conversion from POSS-II to USNO coordinates. 

A few ($\sim10$) objects, out of $\sim300$, are fainter than the limiting
magnitude of the POSS--II, and their positions are therefore accurate to a
few arcsec. 
   
%% on ne montre pas la mosaique finale en IR????

\subsection{Detection \& completeness}

For the tasks of detection and photometry, we applied Sextractor (version 2) 
(Bertin \& Arnouts 1996) to our images and rms map. 
%
%The residual background 
%was removed by setting 64 pixels in RA and 32 pixels in DEC as the mesh for 
%the background determination. 
%
To be detected, objects should have 9 pixels 
brighter than $1\sigma$ on both the filtered (by a Sex standard ``all-ground'' 
convolution mask with $FWHM = 2$ pixels) and the unfiltered image. 

Inside each field, the exposure time and the noise properties do not change 
enough to affect the probability of detection of the objects considered in 
this paper, except near the edge of the field, a region which is hereafter 
excluded from the analysis. The final useful area is $380$ arcmin$^2$; the region
exluded because on the border is $\sim40$ arcmin$^2$.

The completeness was computed in each field as described in Garilli et
al. (1999). In brief, objects are detected when their central surface 
brightness (not their magnitude) is brighter than the detection threshold. 
Objects with a given central surface brighness may have different magnitudes. 
The completeness magnitude at a given central brightness is taken as the 
brightest magnitude of the objects having such central surface brightness (see 
Fig. 4). 

Detected objects in the last half-magnitude bin have typically $S/N\sim5$ 
(see Fig. 5). A note of caution: due to the large number of operations 
performed on the data and the complex way we do it, magnitude errors could be 
non strictly poissonian (or gaussian) in nature, and therefore it could happen 
that the error distribution has broader wings than a gaussian. 

%% fig 5

%%F5

Table 2 shows the completeness magnitudes for several types of magnitudes. 

\begin{table}
\caption{Nominal completness magnitudes}
\begin{tabular}{lccc}
\hline
\small
type & patch1 & 0404 & 0504 \\
\hline
mag$_{iso}$   & 17.2 & 16.9 & 17.2\\
mag$_{kron}$  & 17.2 & 17.0 & 17.2\\
mag$_{5"}$    & 17.4 & 17.2 & 17.4\\
mag$_{10"}$   & 17.1 & 17.0 & 17.2\\
mag$_{13"}$   & 16.9 & 17.8 & 17.9\\
mag$_{14.5"}$ & 16.8 & 16.8 & 16.8\\
mag$_{18.5"}$ & 16.5 & 16.5 & 16.5\\
\hline 
\end{tabular}
\end{table}

\section{Results}

Tables 3, 4 and 5 (given only in electronic form) present the extensive 
photometry in the near-infrared $H$ band of a sample which is complete in all 
the types of magnitudes listed in Table 2, in an area of 
about 400 arcmin$^2$ toward the Coma cluster of galaxies. Figure 1 and 7 
show the studied region and the detected objects. An important point to 
mention is that $H\sim17$ mag is six 
magnitudes below the characteristic magnitude of galaxies, well into the dwarfs' 
regime at the distance of the Coma cluster. To our knowledge, this is the 
first {\it complete} sample of galaxies 
in a nearby cluster for which the photometry is published.

% fig 6

%%F6

Figure 8 shows the classification parameter as a function of $H_{10"}$. 
According to the Sextractor criteria, the classification is secured for most 
objects in this field, at least down to the completeness magnitude. When 
$Class\_star <\sim0.3$, objects are identified as galaxies, whereas $Class\_star 
>\sim 0.7$ (for 0404 and 0504) and $Class\_star >\sim 0.8$ (for patch1) 
correspond to stars. The difference between fields is due to seeing 
conditions. 

Figure 9 displays the result of this classification when comparing 
the isophotal area (at $\mu=22$ mag arcsec$^{-2}$) to the corresponding 
isophotal magnitude. Stars and galaxies easily separate in two different 
sequences in a large magnitude range. This visual criterion is similar to 
the one originally proposed by Godwin \& Peach (1977). The difference between 
the two sequences tends to vanish at faint magnitudes. In our case, the 
separation between stars and galaxies seems to be reliable down the 
completeness limit.

Figure 10 shows a comparison between the different magnitudes obtained in this
field. As expected, magnitudes are sensitive to the size of the apertures 
for bright objects ($H \ltapprox 14$), and apertures larger than $\sim 10"$
give similar results for $H \gtapprox 14$.

Table 3, 4 and 5 give,
for each object brighter than at least one of the limits indicated in Table 2:

-- sky coordinates (epoch J2000)

-- $H$-band magnitude within the apertures: $5'', 10'', 13''$, $14.5''$ and 
$18.5''$, and their errors, as computed by Sex.
Listed errors do not include photometric zero--point errors (see Sect. 3.7),
those arising from the compositing, and implicitly assume a gaussian 
or poissonian error distribution.

-- isophotal ($\mu=22$ mag arcsec$^{-2}$) and Kron magnitudes (with errors). 
Kron magnitudes are computed by integrating the flux in an area 2.5 times larger 
than the Kron area, with a minimun aperture radius of 1.8 arcsec. ``9.99" or 
``99.99" means that no measure is available (usually because
the object is too small
or too faint for its flux to be measured through a large aperture). 

-- isophotal area (at $\mu=22$ mag arcsec$^{-2}$), in arcsec 

-- ellipticity and position angle (from North to East)

-- Kron radius, or 1.8 arcsec if the latter is larger

-- stellarity index as defined by Sextractor

-- an optical identification, usually taken from the Godwin, Metcalfe
\& Peach (1983, and private comunication).

This catalogue is the database used to compute the luminosity function in this 
field (Andreon \& Roser, in preparation), and also to study the
morphology and the photometric properties of these galaxies in details. 

% fig 7 8 e 9

\medskip

{\it Acknowledgements}
We thank Alain Klotz for obtaining our March data the TBL. We also thank 
the TBL technical staff for their support during the observing runs. 
The USNO catalog and the POSS image were obtained at the Canadian Astronomy 
Data Center (CADC), which is operated by the Herzberg Institute of 
Astrophysics, National Research Council of Canada.

\newpage

\begin{figure*}
\epsfysize=16cm
%\centerline{\epsfbox[40 120 545 720]{DS1745.f1}}
\caption[h]{
$H$-band mosaic of the region under investigation. 
Faint objects have $H\sim16$ in this heavly rebinned
and compressed (for display purposis) image.
The field is  $\sim 20 \times 24$ arcmin. North is up 
and East is left. The two dominant galaxies of the Coma cluster are located 
near the South-West corner} 
\end{figure*}

\begin{figure}
\epsfysize=8cm
\centerline{\epsfbox[90 230 525 745]{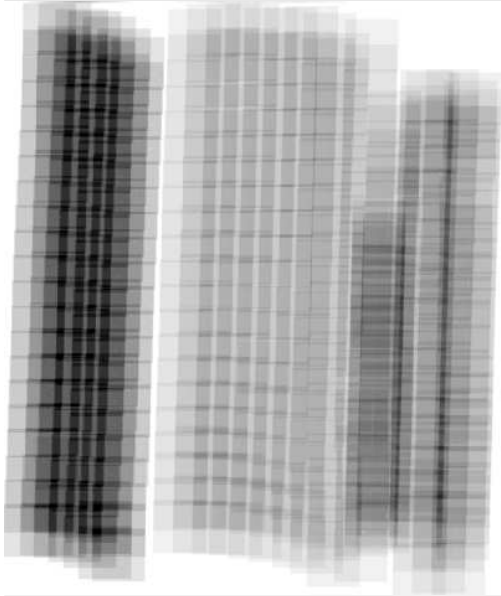}}
\caption[h]{
Exposure map for the whole imaged field ($20\times24$ arcmin wide). North is up and 
East is left. Darker regions were exposed longer. Exposure times range 
from 450 seconds (dark regions), to 0 (white regions). Mean exposure
times are, from left to right, 400, 100 and 310 seconds} 
\end{figure} 

\begin{figure}
\epsfysize=8cm
\centerline{\epsfbox[60 200 460 590]{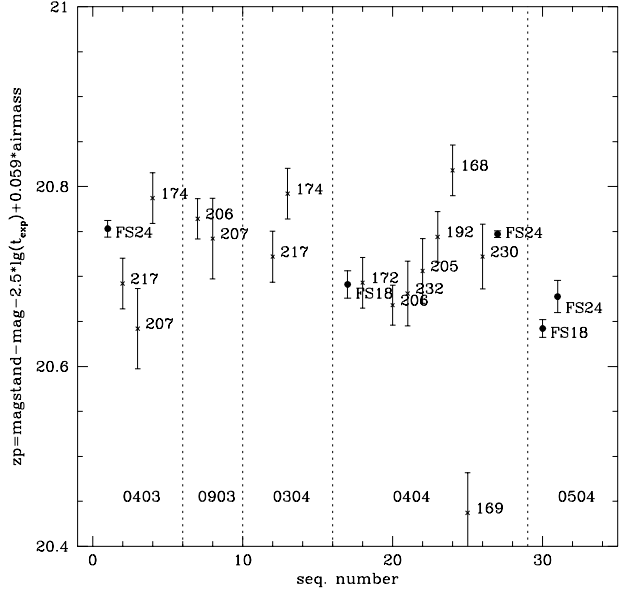}}
\caption[h]{
Magnitude zero point, as measured from standard stars (solid dots) and from 
Coma cluster galaxies (crosses), splitted by date and then sorted by an arbitrary
sequential order. Note that one tickmark along the ordinates 
is 0.05 mag. Error bars for galaxies account for centering error and published 
errors of the standard (statistical errors are virtually null for our measure 
of theses bright objects observed for the photometric calibration), whereas for 
stars they are the ratio of the interquartile range of the various measures by 
the square root of the number of measures.  Labels correspond to names: from 
Dressler (1980) for galaxies and from Casali \& Hawarden (1992) for stars.
} 
\end{figure} 

\begin{figure*}
\epsfysize=15cm
\centerline{\epsfbox[35 170 570 695]{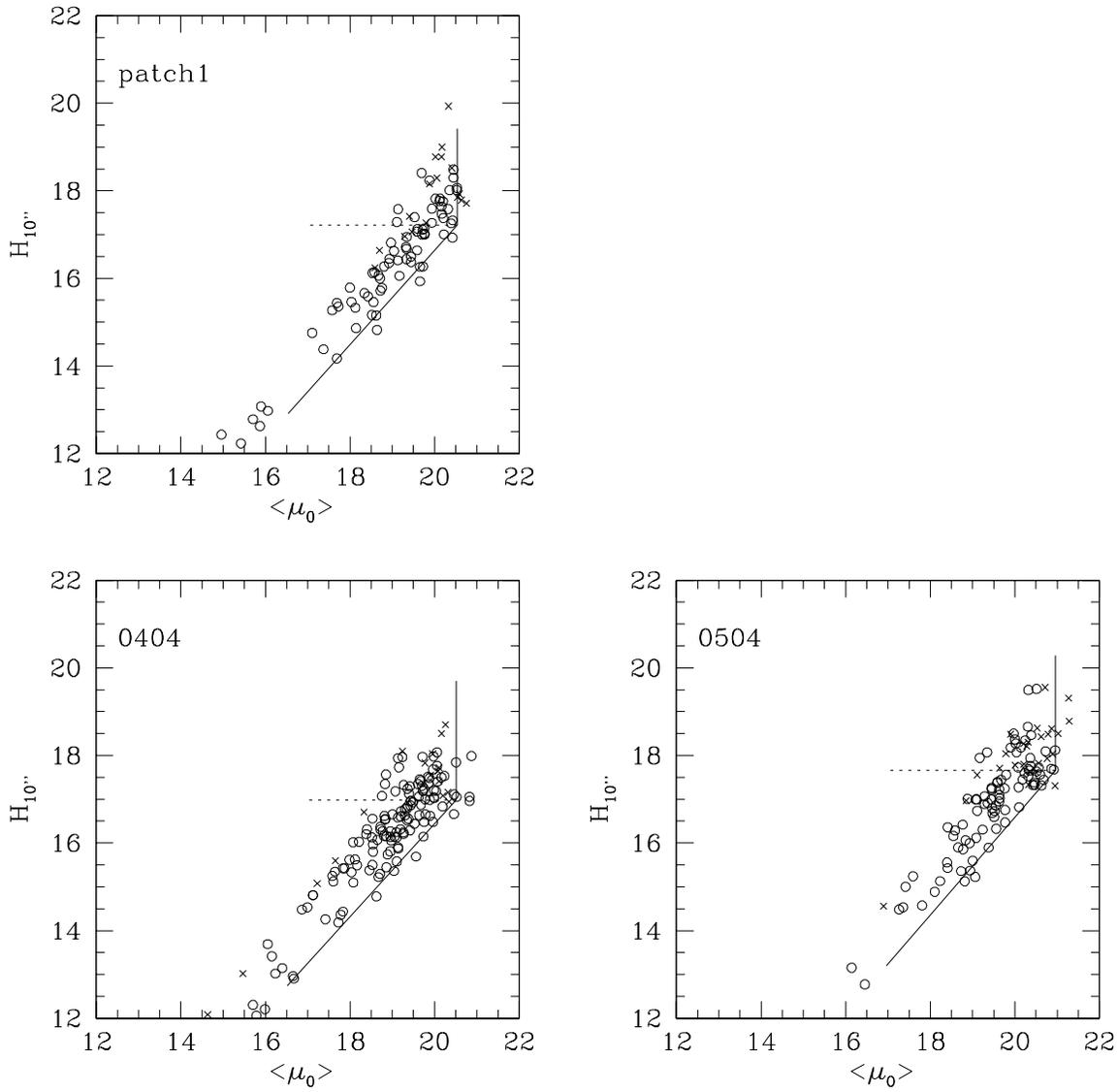}}
\caption[h]{
Average central surface brightness vs. magnitude in a 10\arcsec\ aperture
for the three fields. The completeness magnitude is indicated by the
horizontal dashed line. Circles mark reliable galaxies, crosses objects
with uncertain star/galaxy classification ($0.7<$ stellarity index $<0.8$)}
\end{figure*}

\begin{figure}
\epsfysize=8cm
\centerline{\epsfbox[60 190 470 590]{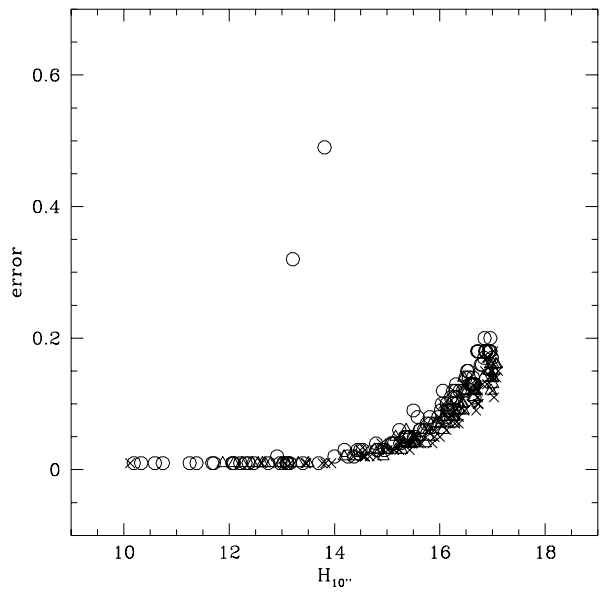}}
\caption[h]{
Magnitude error as a function of the 10\arcsec\ aperture magnitude for patch1 
(triangles), 0404 (circles) and 0504 (squares).  
The two outliers are in a noisier than average region, 
on the border of the region considered for the final catalogue} 
\end{figure}

\begin{figure*}
\epsfysize=15cm
%\centerline{\epsfbox[60 40 550 750]{DS1745.f6}}
\caption[h]{A zoom on a few objects. Faint galaxies in the lower panels
have $H\sim17$}
\end{figure*}

\begin{figure*}
\epsfysize=15cm
\centerline{\epsfbox[45 195 550 675]{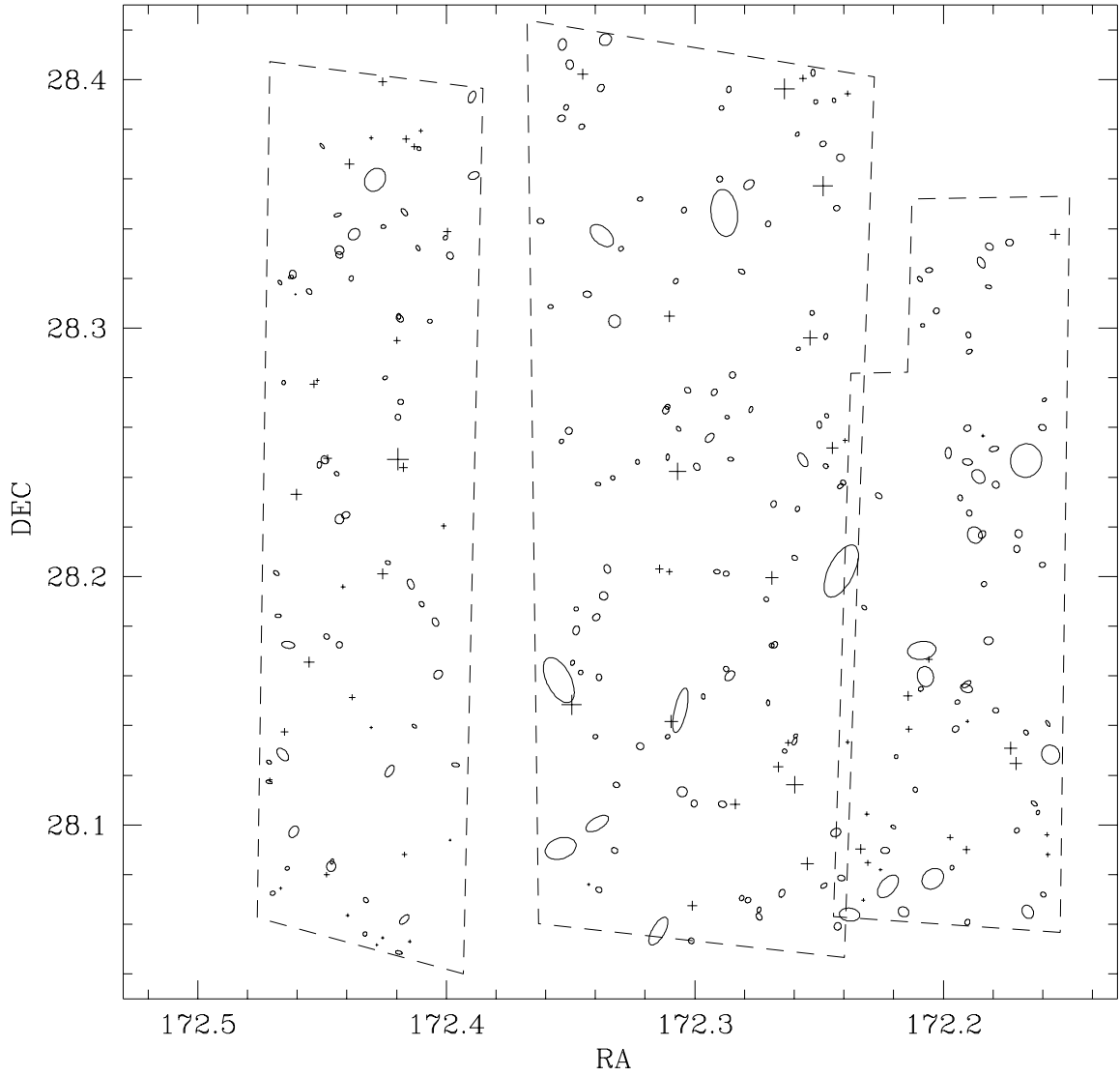}}
\caption[h]{
Detected objects brighter than the completeness magnitude. Crosses are stars, 
ellipses are galaxies. The area of ellipses is twice the 22 mag arcsec$^{-2}$ 
isophotal area. The dashed lines delimit the regions covered by the final catalogue. 
Coordinates are J2000. For simplicity of display, right ascension is given as 
$\alpha*15*\cos 28^{\circ}$} 
\end{figure*}

\begin{figure}
\epsfysize=8cm
\centerline{\epsfbox[60 190 550 675]{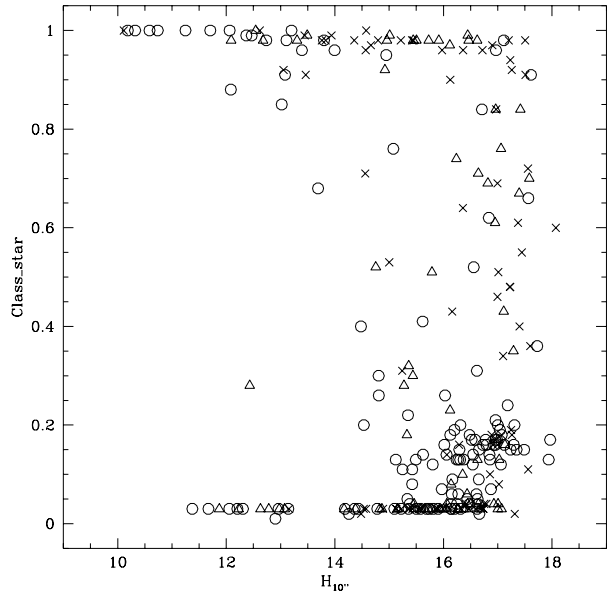}}
\caption[h]{
Classification parameter as a function of the 10\arcsec\ aperture magnitude for
patch1 (triangles), 0404 (circles) and 0504 (crosses)}
\end{figure}

\begin{figure*}
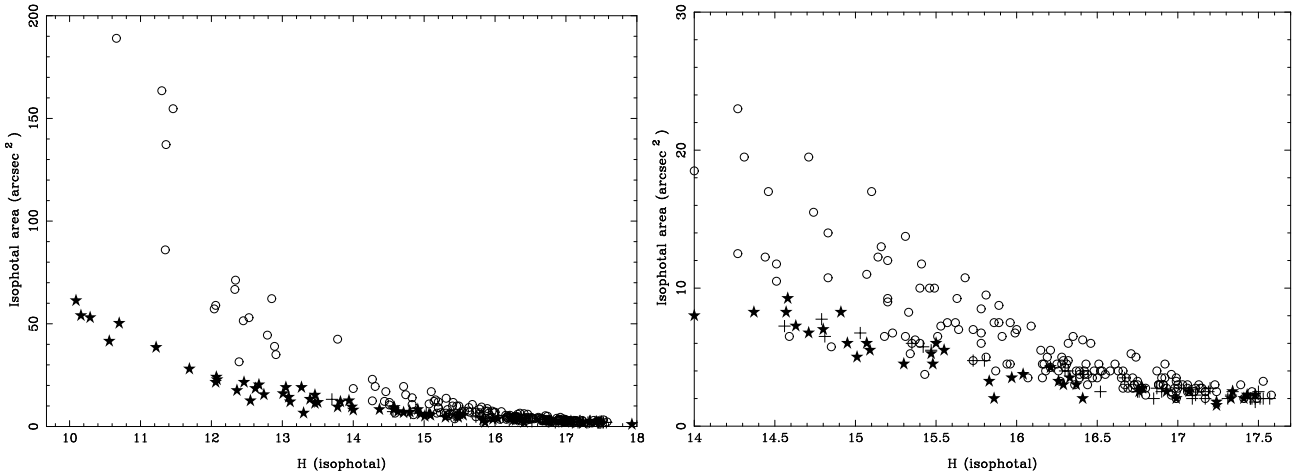

\hbox{
\psfig{file=DS1745.f9a,angle=270,width=8.5cm}
\psfig{file=DS1745.f9b,angle=270,width=8.5cm}
}
\caption{Isophotal area (in arcsec$^2$) versus isophotal magnitude
(corresponding to $\mu=22$ mag arcsec$^{-2}$) in the whole field studied.
Objects identified as stars are shown as black stars,
whereas objects identified as galaxies or unclassified objects
are displayed as open circles and crosses, respectively}
\label{}
\end{figure*}

\begin{figure*}
\epsfysize=15cm
\centerline{\epsfbox[40 200 550 675]{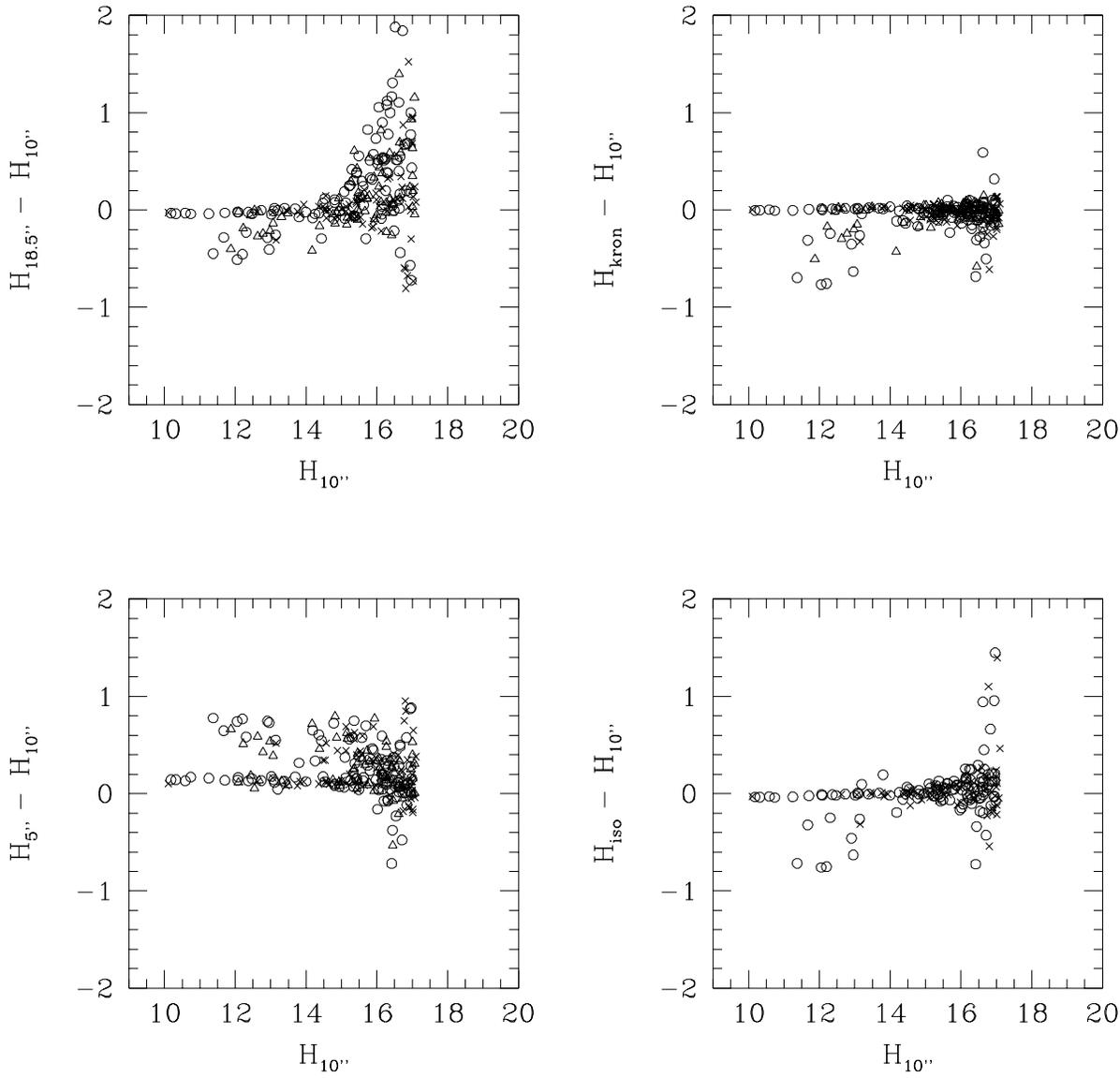}}
\caption[h]{
Residual magnitude vs magnitude diagrams for patch1 (triangles), 
0404 (circles) and 0504 (crosses) for objects brighter than the
$H_{10"}$ completness magnitude. The scatter is due to
variety of surface brightness radial profiles of objects 
in the studied region. Note that Kron and $H_{10"}$ magnitudes
are quite similar at $H_{10"}\sim14$} 
\end{figure*} 


\begin{thebibliography}{}

\bibitem[Bertin \& Arnouts 1996]{1996A&AS..117..393B} Bertin E.,
Arnouts S. 1996, A\&AS 117, 393 

\bibitem[Barger et al. 1996]{}
Barger A.J., 
Aragon-Salamanca A., Ellis R.S., Couch W.J., Smail I., Sharples R.M.,
1996, MNRAS 279, 1 

\bibitem[Butcher Oemler 1984]{} 
Butcher H.R., Oemler A. Jr., 1984, ApJ 285, 426

\bibitem[Casali \& Hawarden 1992]{}
Casali M., Hawarden T., 1992,  JCMT-UKIRT Newsletter n.3, 33

\bibitem[Cowie Songaila Hu \& Cohen 1996]{1996AJ....112..839C}
Cowie, L. L., Songaila, A. , Hu, E. M. \& Cohen, J. G. 1996, AJ 112, 839 

\bibitem[de Lapparent et al. 1997]{} 
de Lapparent V., Galaz G., Arnouts S., Bardelli S., Ramella M., 1997,
The Messenger, 89, 21

\bibitem[De Propris et al. 1998]{}
De Propris R., Eisenhardt P., Stanford S.A., Dickinson M., 1998, ApJ
503, L45

\bibitem[Dressler 1980]{}
Dressler A., 1980, ApJS 42, 565

\bibitem[Gardner et al. 1997]{}
Gardner J.P., Sharples R.M., Frenk C.S., Carrasco B.E., 1997, 
ApJ 480, L99 
 
\bibitem[Garilli et al. 1999]{}
Garilli B., Maccagni D., Andreon S., 1999, A\&A 342, 408

\bibitem[Godwin Metcalfe \& Peach 1983]{1983MNRAS.202..113G} 
Godwin, J. G.,  Metcalfe, N. \& Peach, J. V. 1983, MNRAS 202, 113 

\bibitem[Godwin \& Peach 1977]{}
Godwin J.G., Peach J.V., 1977, MNRAS 181, 323 

\bibitem[Monet et al. 1999]{}
Monet D., Bird A., Canzian B., et al., 1999,
The USN0 -A1 guide star catalog, Washington DC, US Naval Observatory

\bibitem[Pahre et al. 1998]{1998AJ....116.1591P} 
Pahre M.A., Djorgovski S.G., De Carvalho R.R., 1998, AJ, 116, 1591 

\bibitem[Petrosian 1998]{1998ApJ...507....1P} 
Petrosian V., 1998, ApJ, 507, 1 

\bibitem[Recillas-Cruz et al. 1990]{}
Recillas-Cruz E., Carrasco L., Serrano A. et al., 1990, A\&A 229, 64

\bibitem[Saracco et al. 1996]{1996MNRAS.283..865S} 
Saracco P., Chincarini G., Iovino A., 1996, MNRAS, 283, 865 

\bibitem[Szokoly et al. 1998]{}
Szokoly G.P., Subbarao M.U., Connolly A.J., Mobasher B.,  1998,
ApJ 492, 452 
 
\bibitem[Trentham \& Mobasher 1998]{1998MNRAS.299..488T} 
Trentham N., Mobasher B., 1998, MNRAS 299, 488 

\end{thebibliography}
\end{document}